\documentclass[aps,prb,reprint,superscriptaddress,showpacs]{revtex4-1}
\usepackage{graphicx}
\usepackage{bm}

\begin{document}

\title{Third-order nonlinearity by the inverse Faraday effect in planar
magnetoplasmonic structures}
\author{Song-Jin Im}
\email{ryongnam31@yahoo.com or sj.im@ryongnamsan.edu.kp}
\affiliation{Department of Physics, Kim Il Sung University, Taesong
District, Pyongyang, Democratic People's Republic of Korea}
\author{Chol-Song Ri}
\affiliation{Department of Physics, Kim Il Sung University, Taesong
District, Pyongyang, Democratic People's Republic of Korea}
\author{Kum-Song Ho}
\affiliation{Department of Physics, Kim Il Sung University, Taesong
District, Pyongyang, Democratic People's Republic of Korea}
\author{Joachim Herrmann}
\email{jherrman@mbi-berlin.de}
\affiliation{Max-Born-Institute for Nonlinear Optics and Short Pulse Spectroscopy, Max-Born-Str. 2a,
D-12489 Berlin, Germany}
\date{\today}

\begin{abstract}
We predict a new type of ultrafast third-order nonlinearity of surface
plasmon polaritons (SPP) in planar magneto-plasmonic structures caused by
the inverse Faraday effect (IFE). Planar SPPs with a significant
longitudinal component of the electric field act via the IFE as an effective
transverse magnetic field. Its response to the plasmon propagation leads to
strong ultrafast self-action which manifests itself through a third-order
nonlinearity. We derive a general formula and analytical expressions for the
IFE-related nonlinear susceptibility for two specific planar
magneto-plasmonic structures from the Lorentz reciprocity theorem. Our
estimations predict a very large nonlinear third-order nonlinear
susceptibility exceeding those of typical metals such as gold.
\end{abstract}

\pacs{42.65.-k, 73.20.Mf, 75.78.jp, 78.20.Ls}
\keywords{Nonlinear optics, Surface plasmons, Inverse Faraday effect,
Magneto-optical effects}
\maketitle

\affiliation{Department of Physics, Kim Il Sung University, Taesong
District, Pyongyang, Democratic People's Republic of Korea}

\affiliation{Department of Physics, Kim Il Sung University, Taesong
District, Pyongyang, Democratic People's Republic of Korea}

\affiliation{Department of Physics, Kim Il Sung University, Taesong
District, Pyongyang, Democratic People's Republic of Korea}

\affiliation{Max-Born-Institute for Nonlinear Optics and Short Pulse Spectroscopy, Max-Born-Str. 2a,
D-12489 Berlin, Germany}

\section{Introduction}

Nonlinear optical effects can be enhanced by plasmonic structures supporting
local field enhancement and inhomogeneity on the nanoscale \cite%
{Kauranen2012}. In particular, nonlinear propagation of surface plasmon
polaritons (SPPs) in plasmonic waveguides has attracted attention to achieve
ultrafast phase modulation and due to its implication for plasmonic systems%
\cite{MacDonald2008, Mrejen2015}. Different types of nonlinear plasmonic
waveguides has been theoretically investigated, such as nonlinear plasmonic
planar waveguides \cite{Marini2013L, Skryabin2011, Leon2014P, Baron2015},
metal nanowires \cite{Marini2013P, Im2016}, slot waveguides \cite%
{Davoyan2008} and periodic waveguides \cite{Mihalache2010, Suk2014}. In such
plasmonic waveguides, the effective nonlinearity originates from the
third-order optical Kerr effect describing a refractive index change
proportional to the square of the absolute value of the electric field
strength.

Recently, manipulation of the magnetic order of thin magnetic films by
ultrashort pulses based on the inverse Faraday effect (IFE) has attracted
much attention because of its potential impact for future data storage,
spintronics \cite{Kimel2005, Stanciu2007, Radu2011} and improved imaging 
\cite{Malinowski2008, Korff2014, Buttner2015, Korff2015}. Using field
enhancement plasmonic structures has also been studied for the
enhancement of the IFE \cite{Smolyaninov2005prb, Belotelov2012prb,
Hamidi2015} and to achieve control of the magnetization of ferromagnetic
material on the nanoscale \cite{Korff2015}. On the other hand, in the last
years much effort has been devoted to study new ways to control the
properties of surface plasmons using external magnetic fields in
ferromagnetic dielectric \cite{Belotelov2011} or metal \cite{Temnov2016}
layers. In these experiments the magnetization induced by the external
magnetic field leads to a change of the plasmon wavenumber which can be
measured by a plasmonic double-slit interferometer.

In this paper we theoretically predict a new type of ultrafast third-order
nonlinearity of surface plasmon polaritons in planar ferromagnetic plasmonic
structures related with the inverse Faraday effect. In a plasmonic layer,
planar SPP with a longitudinal component of the electric field induce a
magnetization which leads to a third-order nonlinear polarization. This kind
of nonlinearity plays in the plasmon propagation analog effects as the
optical Kerr effect but it originates from a different physical mechanism.
We derive a formula for the IFE-related nonlinear susceptibility and
explicit analytical expressions for two kinds of planar magneto-plasmonic
structures. The IFE-related nonlinear susceptibility differs from the
traditional Kerr-related nonlinear susceptibility by its magnitude,
frequency dependence and its inherent dependence on the material parameters.

\section{IFE-related third-order nonlinear susceptibility for planar
magneto-plasmonic structures}

In a bulk ferromagnetic material, a static external magnetic field leads to
a magnetization $\vec{M}=\vec{M}(\vec{H})$ \cite{Landau1995}.In such material the
vector of the electric displacement depends on the external magnetic
field and is described by

\begin{eqnarray}
\vec{D}={\varepsilon _{0}} \hat \varepsilon \vec{E}={\varepsilon _{0}}\left( {\varepsilon \vec{E}+i\beta \vec{E}\times }%
\vec{M}\right), 
\label{eq1}
\end{eqnarray}
where the relative permittivity tensor $ \hat \varepsilon$ is expressed as
\begin{eqnarray}
\hat \varepsilon  = \left( {\begin{array}{*{20}{c}}
{{\varepsilon}}&i{\beta}M_{z}&-i{\beta}M_{y}\\
-i{\beta}M_{z}&{{\varepsilon}}&i{\beta}M_{x}\\
i{\beta}M_{y}&-i{\beta}M_{x}&{{\varepsilon}}
\end{array}} \right)
\nonumber
\label{eq1_add}
\end{eqnarray}
 and $\beta$ describes the magneto-optical susceptibility. From this
relation one can see that the optical polarization has a contribution
proportional to the magnetization $\vec{M}.$ The consequence of this
relationship is the magneto-optical Faraday effect leading to a polarization
rotation when a linearly polarized light beam is transmitted through a
magneto-optical medium under an external magnetic field. 
The non-diagonal terms in the permittivity tensor describes also the changes to light reflected from a magnetized surface (magneto-optical Kerr effect, MOKE). Pump-probe measurement in nickel films has shown that fast subpicosecond demagnetization can be induced by femtosecond optical pulses \cite{Beaurepaire1996}.

A magnetic material
irradiated by circularly polarized light induces a magnetization along the
wave vector $\vec{k}$ \cite{Pitaevskii1961,vanderZiel1965} which is called the inverse Faraday effect (IFE). The light-induced magnetization can be
expressed as 

\begin{eqnarray}
\vec{M}=-i\chi (\vec{E}\times {\vec{E}^{\ast }})
\label{eq2}
\end{eqnarray}%
where $\chi =$ ${\chi _{g}^{(3)}/\beta}$ \ is a material dependent constant
related with the Verdet constant and ${\chi _{g}^{(3)}}$ can be understood
as the IFE-related third-order nonlinear susceptibility of the bulk
ferromagnetic material. Left- and right-handed polarization waves induce
magnetization of opposite signs.

Light cannot penetrate into a thin metallic layer, but under appropriate
conditions (as e.g. by using the Kretschman configuration for p-polarized
light) surface plasmon-polaritons (SPP) can be excited moving along the
surface. Plasmons exhibit a longitudinal component of the electric field,
therefore the chirality $\vec{E}\times {\vec{E}^{\ast }}$ of plasmons do not
vanishes. This means that even for a linearly polarized input beam a
magnetization can be induced by plasmons, but the polarization of the
plasmon is not circularly polarized.

Substituting the expression (\ref{eq2}) for the magnetization into relation (\ref{eq1}) we
can see that the IFE leads to a third-order nonlinear polarization caused
by a different physical mechanism than the optical Kerr effect. For the
derivation of the IFE-related nonlinear susceptibility of a planar
plasmonic structures including a ferromagnetic layer, we use a formalism
similar as in Ref.~\cite{Im2016}. In a planar waveguides the electromagnetic
field is confined in different spatial modes \cite{Maier2007}, but only the
fundamental mode plays here a role. The mode expansion for the electric and
the magnetic fields $\mathrm{{\vec{F}}}=(\mathrm{{\vec{E}}};\mathrm{{\vec{H}}%
})$ in the plasmonic waveguide can be expressed as $\mathrm{{\vec{F}(}}\vec{r%
},t)=(1/2)\cdot \lbrack \vec{F}(\vec{r})\exp (-i\omega t)+\mathrm{c.c.}]$,
where c.c. signifies the complex conjugate. Below, we restrict ourselves to
the time-independent amplitude for the fundamental mode, $\vec{F}(\vec{r})$,
which can be expressed as 
\begin{eqnarray}
\vec{E}(\vec{r})=\sqrt{1/{s_{0}}}\Psi (x)\exp (i\kappa x){\vec{e}_{0}}({\vec{%
r}_{\bot }}), 
\label{eq3}
\end{eqnarray}
\begin{eqnarray}
\vec{H}(\vec{r}) &=&{(i\omega {\mu _{0}})^{-1}}\nabla \times \vec{E} 
\nonumber \\
&=&\sqrt{1/{s_{0}}}\Psi (x)\exp (i\kappa x){\vec{h}_{0}}({\vec{r}_{\bot }}).
\label{eq4}
\end{eqnarray}
Here $x$ and $\hat{x}$ are the coordinate and the unit vector in the
direction of propagation, ${\vec{r}_{\bot }}$ is the position vector in the
transverse plane, ${s_{0}}$ is defined as ${s_{0}}=(1/2)\int {{%
\mathop{\rm
Re}\nolimits}({{\vec{e}}_{0}}\times \vec{h}_{0}^{\ast })\cdot \hat{x}d\sigma 
}$, where the integral is performed over the transverse plane, $k=\kappa
+i\alpha /2$ is the SPP propagation constant, $\Psi $ is normalized so that
the ${\left\vert \Psi \right\vert ^{2}}$ is equal to the power flow along
the $x$ direction $P(x)=(1/2)\int {{\mathop{\rm Re}\nolimits}(\vec{E}\times {%
{\vec{H}}^{\ast }})\cdot \hat{x}d\sigma }={\left\vert \Psi \right\vert ^{2}}$%
, and ${\vec{e}_{0}}({\vec{r}_{\bot }})$ and ${\vec{h}_{0}}({\vec{r}_{\bot }}%
)$ describe the spatial transverse distribution of the mode. We start from
the Lorentz reciprocity theorem \cite{Snyder1983} 
\begin{eqnarray}
&&\frac{\partial }{{\partial x}}\int {[{{\vec{E}}_{1}}(\vec{r})\times {{\vec{%
H}}_{2}}(\vec{r})-{{\vec{E}}_{2}}(\vec{r})\times {{\vec{H}}_{1}}(\vec{r}%
)]\cdot \vec{x}d\sigma }  \nonumber \\
&=&i\omega \int {\left( {{{\vec{E}}_{1}}(\vec{r})\cdot {{\vec{D}}_{2}}(\vec{r%
})-{{\vec{E}}_{2}}(\vec{r})\cdot {{\vec{D}}_{1}}(\vec{r})}\right) d\sigma },
\label{eq5}
\end{eqnarray}%
where $({\vec{E}_{1}},{\vec{H}_{1}})$ and $({\vec{E}_{2}},{\vec{H}_{2}})$
are two arbitrary guided modes. Now we substitute for $({\vec{E}_{1}},{\vec{H%
}_{1}})$ and $({\vec{E}_{2}},{\vec{H}_{2}})$ the unperturbed backward
propagating field $({\vec{E}_{0}}^{-},{\vec{H}_{0}}^{-})$ and the perturbed
forward propagating field $(\vec{E},\vec{H})$ depending on an external
quasi-static transverse magnetic field $H_{ex}$, correspondingly. The
external magnetic field induces a magnetization $\vec{M}=(0,M,0), M=M(H_{ex})$%
 in the transverse y-direction and leads to a perturbation for the mode
distribution. In the first order of perturbation Eq.~(\ref{eq5}) leads to
the following equation describing the amplitude $\Psi (x)$ of the plasmonic field: 
\begin{eqnarray}
\frac{{d\Psi }}{{dx}}=-\frac{\alpha }{2}\Psi +i\Delta k\cdot \Psi, 
\label{eq6}
\end{eqnarray}%
where $\alpha $ is the linear loss coefficient and%
\begin{eqnarray}
\Delta k=\frac{{i{k_{0}\beta }\int }M{{\cdot {e_{0x}}{e_{0z}}d\sigma }}}{{{%
Z_{0}}\int {({{\vec{e}}_{0}}\times {{\vec{h}}_{0}})\cdot \hat{x}d\sigma }}}
\label{eq7}
\end{eqnarray}%
is the shift of the plasmon wavenumber induced by the external magnetic
field. ${Z_{0}}=\sqrt{{\mu _{0}}/{\varepsilon _{0}}}$ and ${k_{0}}=2\pi
/\lambda $ are the wave impedance and the light wavenumber in vacuum,
respectively.

Now we consider an alternative arrangement without external static magnetic
field but with a laser pulse directed to the planar magneto-optical
structure and acting as an effective magnetic field via the IFE. The SPP
mode of the planar plasmonic waveguide induces a magnetization into the
transverse y-direction. From Eq.~(\ref{eq2}) the nonlinear magnetization $%
\vec{M}=(0,M,0)$ is expressed as%
\begin{eqnarray}
M=-i\chi (E_{x}^{\ast }{E_{z}}-{E_{x}}E_{z}^{\ast })
\label{eq8}
\end{eqnarray}%
with $\chi ={\chi _{g}^{(3)}/\beta }$. If we substitute Eq.~(\ref%
{eq8}) into Eq.~(\ref{eq6}) and (\ref{eq7}), we find 
\begin{eqnarray}
\frac{{d\Psi }}{{dx}}=-\frac{\alpha }{2}\Psi +i\gamma {\left\vert \Psi
\right\vert ^{2}}\Psi, 
\label{eq9}
\end{eqnarray}%
where%
\begin{eqnarray}
\gamma ={k_{0}}\frac{{2\int \chi _{g}^{(3)}{{e_{0x}}{e_{0z}}\left( {%
e_{0x}^{\ast }{e_{0z}}-{e_{0x}}e_{0z}^{\ast }}\right) d\sigma }}}{{{Z_{0}}%
\int {({{\vec{e}}_{0}}\times {{\vec{h}}_{0}})\cdot \hat{x}d\sigma }\cdot
\int {{\mathop{\rm Re}\nolimits}({{\vec{e}}_{0}}\times \vec{h}_{0}^{\ast
})\cdot \hat{x}d\sigma }}}, 
\label{eq10}
\end{eqnarray}%
is the effective nonlinear propagation coefficient.

\section{Ferromagnetic dielectric/metal interface}

\begin{figure}[tbp]
\includegraphics[width=0.5\textwidth]{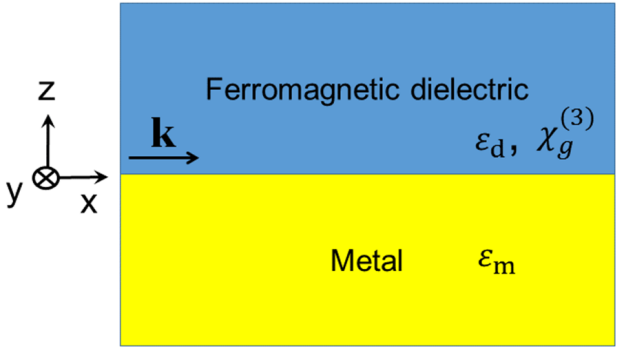}
\caption{The ferromagnetic dielectric/metal interface.}
\label{fig:1}
\end{figure}

\begin{figure}[tbp]
\includegraphics[width=0.5\textwidth]{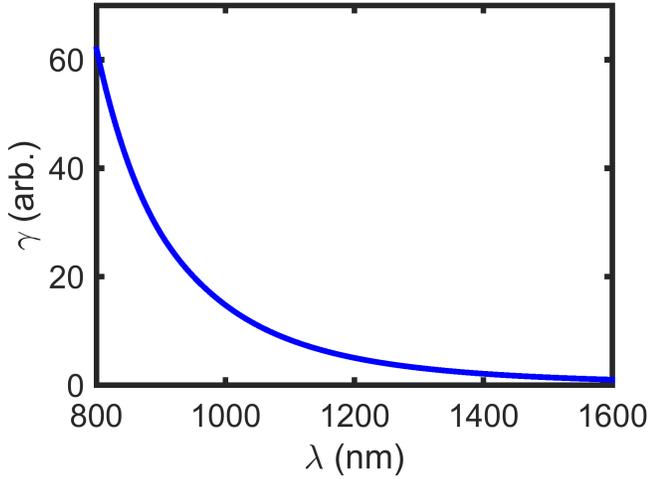}
\caption{Wavelength-dependence of the IFE-related nonlinear susceptibility
for the interface between gold and a ferromagnetic dielectric. The nonlinear
susceptibility of the bulk ferromagnetic dielectric $\protect\chi _g^{(3)}$
is assumed to be independent on the wavelength.}
\label{fig:2}
\end{figure}

\begin{figure}[tbp]
\includegraphics[width=0.5\textwidth]{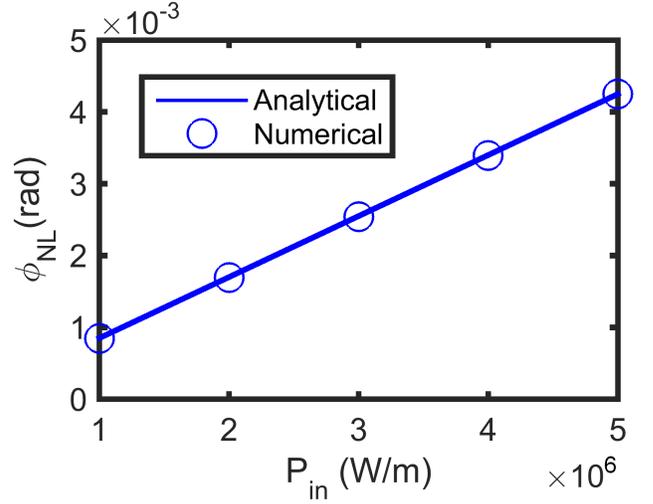}
\caption{Power-dependence of the nonlinear phase shift for a wavelength of
1550 nm and a propagation distance of $L$=1000 nm for the interface between
gold and a ferromagnetic dielectric with ${\protect\varepsilon _d}
= 4$. The blue line represents ${\protect\phi_{%
\mathrm{NL}}} = \protect\gamma PL$ using the analytical formula of $\protect%
\gamma$ by Eq.~(\protect\ref{eq12}). The blue circles represent ${\protect%
\phi_{\mathrm{NL}}}(P) = \protect\phi ({P_{\mathrm{in}}} = P) - \protect\phi %
({P_{\mathrm{in}}} \to 0)$ calculated by the numerical solutions of Maxwell
equations in the frequency domain. Here we assumed $\protect\chi _g^{(3)} = {%
10^{ - 17}}{\mathrm{m}^2}{\mathrm{V}^{ - 2}}$.}
\label{fig:3}
\end{figure}

Let us derive more explicit analytical formulas for two typical types of
planar magneto-plasmonic interfaces. First, we consider a ferromagnetic
dielectric/metallic interface as shown in the Fig.~\ref{fig:1}.

In the case that an external magnetic field $H_{ex}$ is present leading to a
magnetization $M=M(H_{ex})$, by substituting the analytical expression of the
fundamental TM mode distribution of a single interface Eq. (2.10)-(2.14) of Ref.~\cite{Maier2007}
into Eq.~(\ref{eq7}) we can derive a plasmon wavenumber shift given by 
\begin{eqnarray}
\Delta k={k_{0}}\frac{\beta M}{{\sqrt{-({\varepsilon _{m}}+{\varepsilon _{d}}%
)}(1-\varepsilon _{d}^{2}/\varepsilon _{m}^{2})}}. 
\label{eq11}
\end{eqnarray}%
Eq.~(\ref{eq11}) is in agreement with Eq.~(4) of Ref.~\cite{Belotelov2011}
used for the description of the control of the optical phase of a plasmon in
a magneto-optical interferometer.

For the case without external magnetic field but with an incident laser
pulse the substitution of the analytical mode distribution for the
fundamental mode of the planar plasmonic waveguide \cite{Maier2007} into
Eq.~(\ref{eq10}) yields the following expression: 
\begin{eqnarray}
\gamma \approx \frac{{\ -4\chi _{g}^{(3)}}}{{c{\varepsilon _{0}}}}\frac{{%
\varepsilon _{m}^{4}{\varepsilon _{d}}}}{{{{\left( {\varepsilon
_{m}^{2}-\varepsilon _{d}^{2}}\right) }^{2}}\left( {{\varepsilon _{m}}+{%
\varepsilon _{d}}}\right) }}\sqrt{\frac{{\ -1}}{{{\varepsilon _{m}}+{%
\varepsilon _{d}}}}}k_{0}^{2}. 
\label{eq12}
\end{eqnarray}%
Here, high quality of the plasmonic metal was assumed so that ${%
\mathop{\rm
Im}\nolimits}{\varepsilon _{m}}<<\left\vert {{\mathop{\rm Re}\nolimits}{%
\varepsilon _{m}}}\right\vert $.

Fig.~\ref{fig:2} shows the wavelength-dependence of the nonlinear
coefficient $\gamma $. Here, we used the frequency-depending experimental
data for the permittivity of gold \cite{Johnson1972} as ${\varepsilon _{m}}%
(\lambda )$. The wavelength-dependence of the nonlinear susceptibility of
the bulk ferromagnetic dielectric $\chi _{g}^{(3)}$ is disregarded.

Fig.~\ref{fig:3} shows the power-dependence of the nonlinear phase shift ${%
\phi _{\mathrm{NL}}}$ for a wavelength of 1550 nm and a propagation distance
of $L=1000$ nm for an interface between gold and a ferromagnetic dielectric.
For the calculation of the absolute values of the nonlinear phase shift, we
assumed $\chi _{g}^{(3)}={10^{-17}}{\mathrm{m}^{2}}{\mathrm{V}^{-2}}$ and ${%
\varepsilon _{d}}=4$. The blue line represents ${\phi _{\mathrm{NL}}}=\gamma
PL$ with the analytical prediction of $\gamma $ by Eq.~(\ref{eq12}). The
analytical prediction is in good agreement with the numerically determined
phase shift ${\phi _{\mathrm{NL}}}(P)=\phi ({P_{\mathrm{in}}}=P)-\phi ({P_{%
\mathrm{in}}}\rightarrow 0)$ (the blue circles in Fig.~\ref{fig:4}),
obtained by numerical solutions of the Maxwell equations in the frequency
domain (analogous as in Ref.~ \cite{Im2016}).

If we consider ${\varepsilon _{d}}<<\left\vert {\varepsilon _{m}}\right\vert 
$ in the infrared region, Eq.~(\ref{eq12}) is simplified to 
\begin{eqnarray}
\gamma \approx \frac{{4\chi _{g}^{(3)}k_{0}^{2}}}{{c{\varepsilon _{0}}}}%
\frac{{\varepsilon _{d}}}{{{{(-{\varepsilon _{m}})}^{3/2}}}}. 
\label{eq13}
\end{eqnarray}%
The IFE-related nonlinear susceptibility is predicted to be linearly
dependent on the permittivity of the dielectric material as can be seen in
Eq.~(\ref{eq13}). Note that for the same material besides the IFE-related
third-order nonlinearity a different type of third-order nonlinearity exist
caused by the optical Kerr effect. Substituting the mode distribution of the
planar interface to Eq.~(9) of Ref.~\cite{Im2016}, the Kerr-related
nonlinear susceptibility of the planar interface is given by 
\begin{eqnarray}
\gamma \approx \frac{{3\chi _{k}^{(3)}k_{0}^{2}}}{{4c{\varepsilon _{0}}}}%
\frac{1}{{{{(-{\varepsilon _{m}})}^{1/2}}}},
\label{eq14}
\end{eqnarray}%
which is independent on the permittivity of the dielectric $\varepsilon _{d}$. ${\chi
_{k}^{(3)}}$ is here the Kerr-related nonlinear susceptibility of the bulk
dielectric.

\section{Dielectric/hybrid metal-ferromagnetic interface}

\begin{figure}[tbp]
\includegraphics[width=0.5\textwidth]{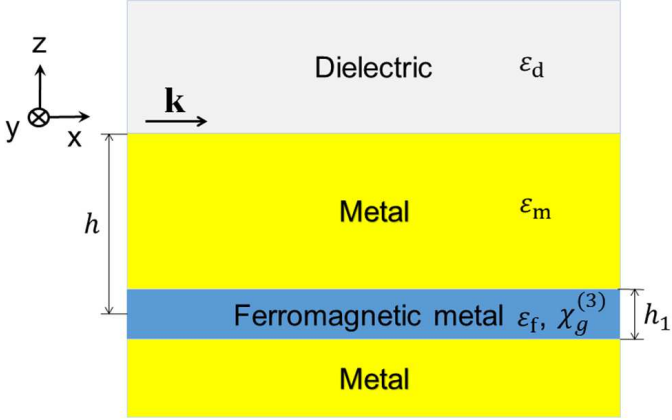}
\caption{The dielectric/hybrid metal-ferromagnet interface.}
\label{fig:4}
\end{figure}

\begin{figure}[tbp]
\includegraphics[width=0.5\textwidth]{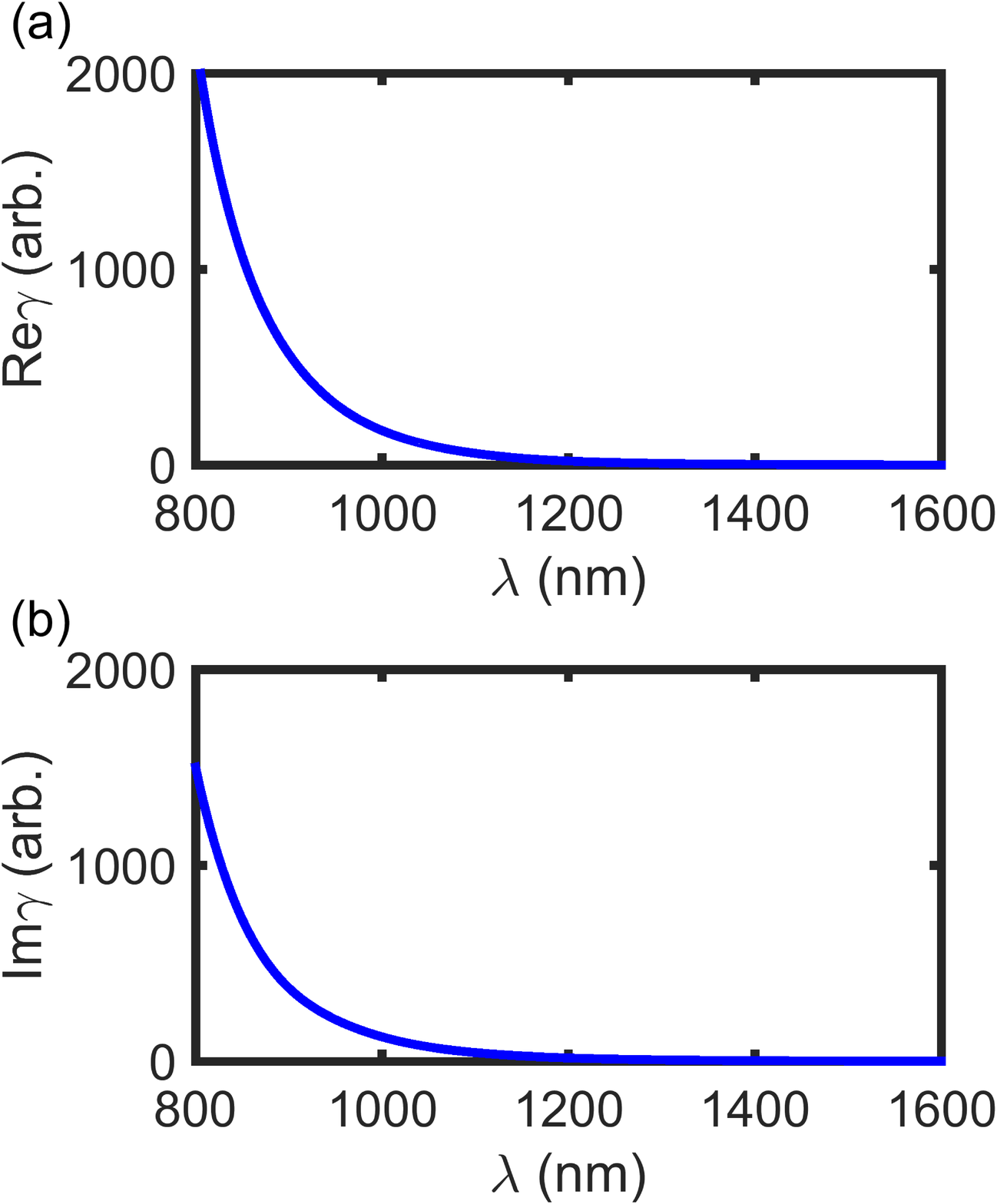}
\caption{Wavelength-dependence of the IFE-related nonlinear susceptibility
for the interface between the Au-Co-Au hybrid structure and air. The
wavelength-dependence of $\protect\chi _g^{(3)}$ is disregarded. (a) and (b)
show the real and the imaginary parts of the nonlinear susceptibility,
respectively.}
\label{fig:5}
\end{figure}

\begin{figure}[tbp]
\includegraphics[width=0.5\textwidth]{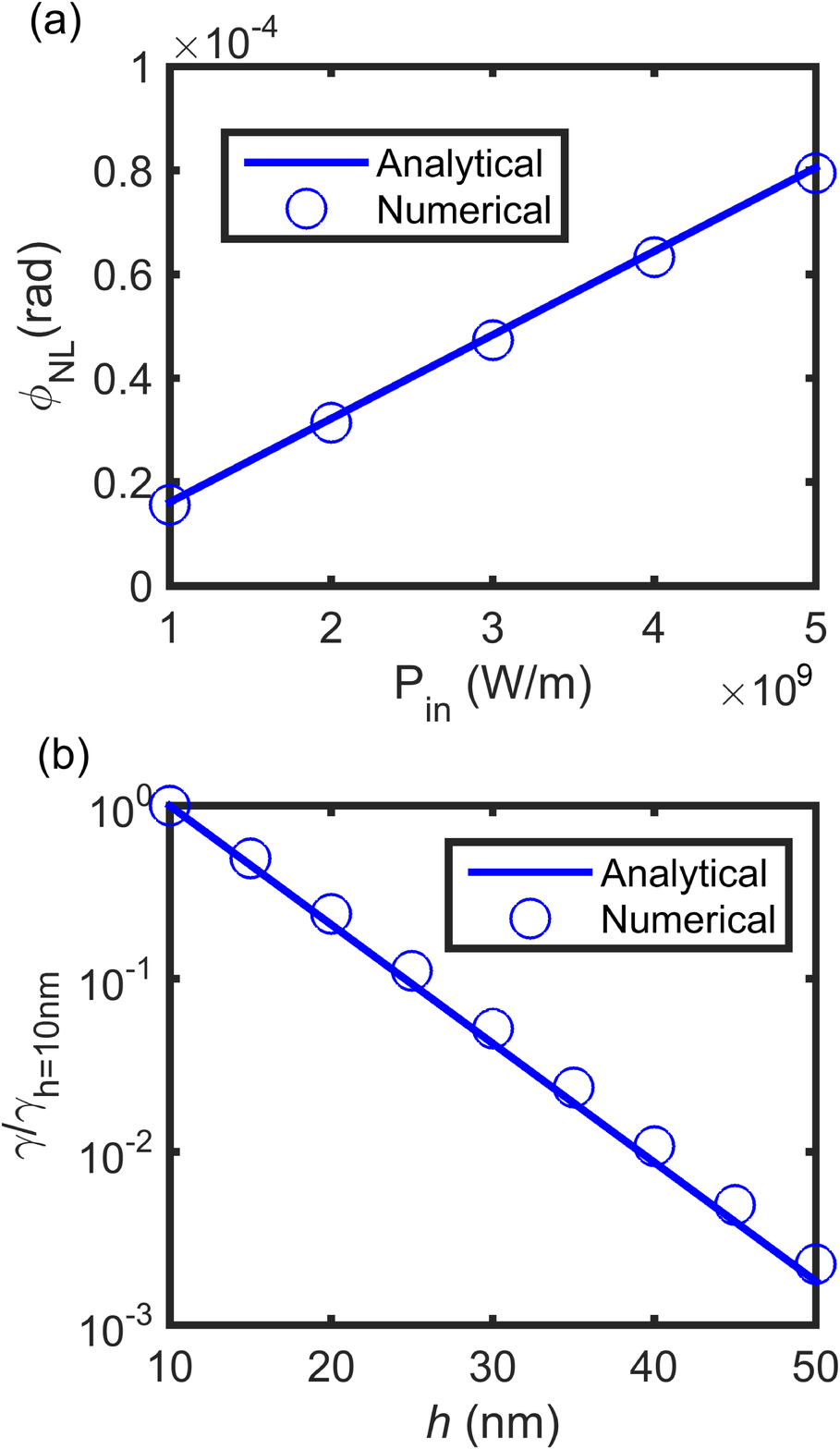}
\caption{(a) Power-dependence of the nonlinear phase shift for a wavelength
of 808 nm and a propagation distance of $L$=5000 nm for the interface
between the Au-Co-Au hybrid structure and air with $\varepsilon _{d}=1$, $h=20$ nm and $h_{1}=4$ nm. (b) Dependence of the
nonlinear susceptibility on the position of the cobalt layer. Here, the blue
line is calculated by the analytical expression from Eq.~(\protect\ref{eq16}%
) and the blue circles are by numerically solving the Maxwell equations in
the frequency domain. Here we assumed $\protect\chi _g^{(3)} = {10^{ - 17}}{%
\mathrm{m}^2}{\mathrm{V}^{ - 2}}$.}
\label{fig:6}
\end{figure}

Next, we consider the dielectric/hybrid metal-ferromagnetic interface as
shown in the Fig.~\ref{fig:4}. In particular such structure has been applied
for active magneto-plasmonic micro-interferometry \cite{Temnov2010}.

By substituting the mode distribution of a single interface \cite{Maier2007}%
 into Eq.~(\ref{eq7}) for the case of the presence of an external magnetic
field as in \cite{Temnov2010} we obtain for the plasmon wavenumber shift 
\begin{eqnarray}
\Delta k=-\frac{g}{{\varepsilon _{f}}}\frac{{2{h_{1}}{{({k_{0}}{\varepsilon
_{d}})}^{2}}}}{{({\varepsilon _{d}}+{\varepsilon _{m}})(1-\varepsilon
_{d}^{2}/\varepsilon _{m}^{2})}}\exp (-2{k_{m}}h).
\label{eq15}
\end{eqnarray}%
Here, ${\varepsilon _{f}}$, $h$ and ${h_{1}}$ is the diagonal permittivity,
depth and thickness of the ferromagnetic material, respectively, and ${k_{m}}%
={({k^{2}}-k_{0}^{2}{\varepsilon _{m}})^{1/2}}$. The Eq.~(\ref{eq15})
coincides to Eq.~(2) of Ref.~\cite{Temnov2010}.

Next we consider the case that an incident laser field induces an
opto-magnetic field by the IFE. If we substitute the mode distribution
to Eq.~(\ref{eq10}) and assume ${\mathop{\rm Im}\nolimits}{\varepsilon _{m}}%
<<\left\vert {{\mathop{\rm Re}\nolimits}{\varepsilon _{m}}}\right\vert $, 
\begin{eqnarray}
\gamma &\approx &\frac{{16}}{{c{\varepsilon _{0}}}}\chi _{g}^{(3)}\frac{{{%
\mathop{\rm Re}\nolimits}({\varepsilon _{f}})}}{{\varepsilon _{f}}}\frac{{%
\varepsilon _{m}^{4}\varepsilon _{d}^{4}}}{{{{\left\vert {\varepsilon _{f}}%
\right\vert }^{\mathrm{2}}}{{\left( {\varepsilon _{d}^{2}-\varepsilon
_{m}^{2}}\right) }^{2}}{{\left( {{\varepsilon _{m}}+{\varepsilon _{d}}}%
\right) }^{2}}}}\cdot  \nonumber \\
&&k_{0}^{3}{h_{1}}\exp (-4{k_{m}}h).  
\label{eq16}
\end{eqnarray}%
Fig.~\ref{fig:5} shows the wavelength-dependence of the nonlinear
coefficient for the interface between the Au-Co-Au hybrid structure and air.
Here, we used the experimental data for the permittivity spectra of gold \cite{Johnson1972}  and cobalt 
\cite{Johnson1974} as ${\varepsilon _{m}}(\lambda )$ and ${\varepsilon _{f}}(\lambda )$, respectively. Fig.~\ref{fig:6}(a)
shows the power-dependence of the nonlinear phase shift ${\phi _{\mathrm{NL}}%
}$ for a wavelength of 808 nm and a propagation distance of $L$=5000 nm for
the interface between the Au-Co-Au hybrid structure and air. For calculation
of the absolute values of the nonlinear phase shift, we assumed $\chi
_{g}^{(3)}={10^{-17}}{\mathrm{m}^{2}}{\mathrm{V}^{-2}}$. The analytical
prediction (the blue line) is in good agreement with the numerical
simulation (blue circles). Fig.~\ref{fig:6}(b) shows the exponential
dependence of the nonlinear susceptibility on the position of the cobalt
layer in agreement with the numerical simulation, which demonstrates that
the nonlinearity originates from the thin cobalt layer.

If we consider ${\varepsilon _{d}}<<\left\vert {\varepsilon _{m}}\right\vert 
$ in the infrared region, Eq.~(\ref{eq16}) is simplified to 
\begin{eqnarray}
\gamma \approx \frac{{16}}{{c{\varepsilon _{0}}}}\chi _{g}^{(3)}\frac{{{%
\mathop{\rm Re}\nolimits}({\varepsilon _{f}})}}{{\varepsilon _{f}}}\frac{{%
\varepsilon _{d}^{4}}}{{{{\left\vert {\varepsilon _{f}}\right\vert }^{2}}%
\varepsilon _{m}^{2}}}k_{0}^{3}{h_{1}}\exp (-4{k_{m}}h).  
\label{eq17}
\end{eqnarray}%
As seen the IFE-related nonlinear susceptibility depends on the ${4^{\mathrm{%
th}}}$-power of the permittivity of the dielectric material. For example, by exchanging the air $\varepsilon _{d}=1$ with the garnet $\varepsilon _{d}=6$ in the hybrid structure Fig.~\ref{fig:4}, the IFE-related nonlinear susceptibility is enhanced by more than 1000 times. In Ref.~\cite{ Becerra2012} a possibility has been discussed to increase the magneto-optical effect by increasing $\varepsilon_{d}$, but the drawback of this approach is the simultaneous reduction of the SPP propagation length.
We note, however, the 4th-power-depenedence of the IFE-related nonlinear susceptibility on $\varepsilon_{d}$ dominates the reduction of the SPP propagation length.
Substituting the mode distribution to Eq.~(9)
of Ref.~\cite{Im2016}, we find for the Kerr-related nonlinear
susceptibilities of the structure in Fig.~\ref{fig:4}
\begin{eqnarray}
\gamma \approx \frac{{3\chi _{k}^{(3)}}}{{c{\varepsilon _{0}}}}\frac{{%
\varepsilon _{d}^{3}}}{{\varepsilon _{m}^{3}}}k_{0}^{3}{h_{1}}\exp (-4{k_{m}}%
h),
\label{eq18}
\end{eqnarray}%
where $\chi _{k}^{(3)}$ is the Kerr-related nonlinear  susceptibility in the thin layer with the thickness $h_1$.

Let us compare the magnitude of the nonlinear phase shift of the hybrid structure of Fig.~\ref{fig:4} (including a ferromagnetic metallic layer) with the ferromagnetic dielectric/metal interface in Fig.~\ref{fig:1}. As seen from the comparison of Fig.~\ref{fig:3} with Fig.~\ref{fig:6}(a) the nonlinear phase shift in the ferromagnetic dielectric/metal interface is by orders of magnitude larger. The reason is that in a ferromagnetic dielectric much more energy is distributed than in a metallic structure. 

Note that recently magnetization-induced second harmonic generation has been studied in Ref.~\cite{Zheng2015,Razdolski2016} arising in magneto-plasmonic systems in the presence of an external magnetic field. The here studied third-order IFE-related nonlinear effect qualitatively differ from this second-order nonlinear effect and do not require any external magnetic field.

Let us discuss the time-response of the IFE-related nonlinearity based on the magnetization dynamics in ferromagnetic thin films. The relaxation process of electrons and spin systems in a ferromagnetic thin films after excitation with a femtosecond pulse is related with a number of processes in the interaction of light with the spin degrees of freedom of electrons and the thermalization of electron in such system. In Ref.~\cite{Beaurepaire1996} it was found that a nickel thin film can be demagnetized after excitation by a sub-100fs laser pulse. Several studies confirmed later this result. Note that the underlying mechanism in time-domain magnetization dynamics is still in discussion \cite{Zhang2009,Carva2011}.

\section{Discussions and Conclusions}

Let us estimate the order of magnitudes of parameters for the IFE-related
third-order nonlinearity of SPPs in planar magneto-plasmonic structures.
From the expression for the IFE-induced effective magnetic field ${\vec{H}_{%
\mathrm{eff}}}=-i{\varepsilon _{0}}/{\mu _{0}}\beta (\vec{E}\times {\vec{E}%
^{\ast }})$ we can predict that the nonlinear susceptibility of a
bulk-ferromagnetic material $\chi _{g}^{(3)}$ is on the order of ${%
\varepsilon _{0}}/{\mu _{0}}\cdot {\beta ^{2}}$, where $\beta $ is the
magneto-optical susceptibility. If $\beta $ is on the order of ${10^{-6}}%
\mathrm{m/A}$ \cite{Vahaplar2012} for a ferromagnetic material, $\chi
_{g}^{(3)}$ is on the order of $\ {10^{-17}}{\mathrm{m}^{2}}{\mathrm{V}^{-2}}
$ . This presents a very strong third-order nonlinear susceptibility
compared to the Kerr-nonlinearity of typical dielectric materials with $\chi
_{k}^{(3)}$ on the order of ${10^{-22}}{\mathrm{m}^{2}}{\mathrm{V}^{-2}}$. $%
\chi _{g}^{(3)}$ is also larger than the measured $\chi _{k}^{(3)}$ of gold
on the order of ${10^{-19}}{\mathrm{m}^{2}}{\mathrm{V}^{-2}}$ at the
wavelengths of 630 nm \cite{Rotenberg2007} and 796.5 nm \cite{Leon2014O}.
If we assume a structure depth on the order of a half wavelength which does not induce noticeable deterioration to the device performance \cite{Cai2009}, from Eq.~(\ref{eq12}) we can estimate a huge nonlinear coefficient $\gamma$ on the order of ${10^6}{{\rm{W}}^{ - 1}}{{\rm{m}}^{ - 1}}$.

In conclusion, in this paper we predicted a new type of ultrafast third--order nonlinearity
of SPPs in planar magneto-plasmonic structures based on the induced
effective magnetic field by the inverse Faraday effect and its response on
the plasmon propagation. We derived a formula for the IFE-related nonlinear
susceptibility for two planar magneto-plasmonic structures from the Lorentz
reciprocity theorem and analytical expressions for the nonlinear
coefficients that describe a strong self-action of the SPPs 
manifesting in a nonlinear phase shift and a self-induced absorption. Our
theoretical prediction of the IFE-related nonlinearity indicates a very
large, ultrafast effective third-order susceptibility exceeding those of
typical metals like gold. The results presented here could have important
implication for the study of magneto-plasmonic systems as well as for
applications in nonlinear plasmonics as e.g. to achieve ultrafast plasmonic
modulation.

\bibliography{nonlinearity}

\end{document}